\documentclass{emulateapj}

\usepackage{graphics}

\def\msun{${\rm M_{\odot}}$}
\def\Msun{${\rm M_{\odot}} \;$}
\def\be{\begin{equation}}
\def\ee{\end{equation}}
\def\gcc{g cm$^{-3}$}

\def\bi{\begin{itemize}}
\def\ei{\end{itemize}}

\def\ben{\begin{enumerate}}
\def\een{\end{enumerate}}

\def\bea{\begin{eqnarray}}
\def\eea{\end{eqnarray}}

\begin{document}

\shorttitle{Black Hole White Dwarf Encounters}

\shortauthors{Rosswog, Ramirez-Ruiz, Hix \& Dan}

\title{Simulating Black Hole White Dwarf Encounters}

\author{Stephan Rosswog\altaffilmark{1}, Enrico
  Ramirez-Ruiz\altaffilmark{2},  William R. Hix\altaffilmark{3}, Marius Dan\altaffilmark{1}}
\altaffiltext{1}{School of Engineering and Science, Jacobs University
  Bremen, Campus Ring 1, 28759 Bremen, Germany}
\altaffiltext{2}{Department of Astronomy and Astrophysics, University
  of California, Santa Cruz, CA 95064}
\altaffiltext{3}{Physics Division, Oak Ridge National Laboratory, Oak Ridge,
  TN37831-6374}

\begin{abstract}
The existence of supermassive black holes lurking in the centers of galaxies 
and of stellar binary systems containing a black hole with a few solar masses
has been established beyond reasonable doubt. The idea that black holes of
intermediate masses ($\sim 1000$ \msun) may exist in globular star clusters
has gained credence over recent years but no conclusive evidence has been
established yet. An attractive feature of this hypothesis is the
potential to not only disrupt solar-type stars but also compact white dwarf
stars. In close encounters the white dwarfs can be sufficiently compressed 
to thermonuclearly explode. The detection of an underluminous thermonuclear
explosion accompanied by a soft, transient X-ray signal would be compelling
evidence for the presence of intermediate mass black holes in stellar
clusters.   
In this paper we focus on the numerical techniques used to simulate the entire
disruption process from the initial parabolic orbit, over the  nuclear energy
release during tidal compression, the subsequent ejection of freshly
synthesized material and the formation process of an accretion disk around the
black hole. 
\end{abstract}


\section{Introduction}
\label{intro}
The existence of two classes of black holes is well-established: supermassive
black holes with masses beyond $10^6$ M$_\odot$ are thought to lurk in the
centers of most galaxies\cite{richstone98} and black holes with just a few 
solar masses have been identified as the unseen components in X-ray binary
systems\cite{lewin06}. Black holes with $M_{\rm bh}\sim 1000$ M$_\odot$,
so-called ``intermediate mass black holes'', represent a plausible, but so far
still unconfirmed ``missing link'' between these two well-established classes 
of black holes. 
In recent years, the stellar dynamics in the centers of some globular clusters
has been interpreted as being the result the gravitational interaction with an
intermediate mass black
hole\cite{gebhardt02,gerssen02,gerssen03,gebhardt05}. Further circumstantial
evidence comes from  ultraluminous, compact X-ray sources in young star
clusters\cite{zezas02,pooley06} and from n-body simulations\cite{portegies04}
that indicate that runaway collisions in dense
young star clusters can lead to rapidly growing black holes. None of these
arguments is conclusive by itself, but their different nature suggests that
the possibility of intermediate mass black holes must be taken seriously.\\ 
The tidal disruption of white dwarfs offers the unique possibility to explore
the presence of an intermediate mass black hole. The corresponding disruption
processes have been explored with various approximations in earlier studies
\cite{luminet89b,wilson04,dearborn05}, our simulations for the first time
explore the full evolution from the initial parabolic orbit over the
disruption process to the subsequent build-up of an accretion disk.
We have performed a large set of calculations to identify observational
signatures that can corroborate or, alternatively, rule out the existence of
intermediate mass black holes. In this paper we focus on the numerical
techniques that have been employed in this disruption study, a detailed
discussion of the astrophysical implications will be given
elsewhere\cite{rosswog07f,rosswog07d}.

\section{Numerical methods}
\label{num_meth}
The simulation of a white dwarf disruption by a black hole needs to follow the
gas dynamics from the initial spherical star through the distortion and
compression while approaching the black hole to the subsequent expansion phase
and the formation of an accretion disk. Of paramount importance for the
dynamical evolution is the inclusion of the feedback from the nuclear
reactions that are triggered by the tidal compression. In the following we
will briefly sketch the methods employed in our simulations.

\subsection{Hydrodynamics}
\begin{figure}[htbp] 
   \centering
   \includegraphics[width=14cm]{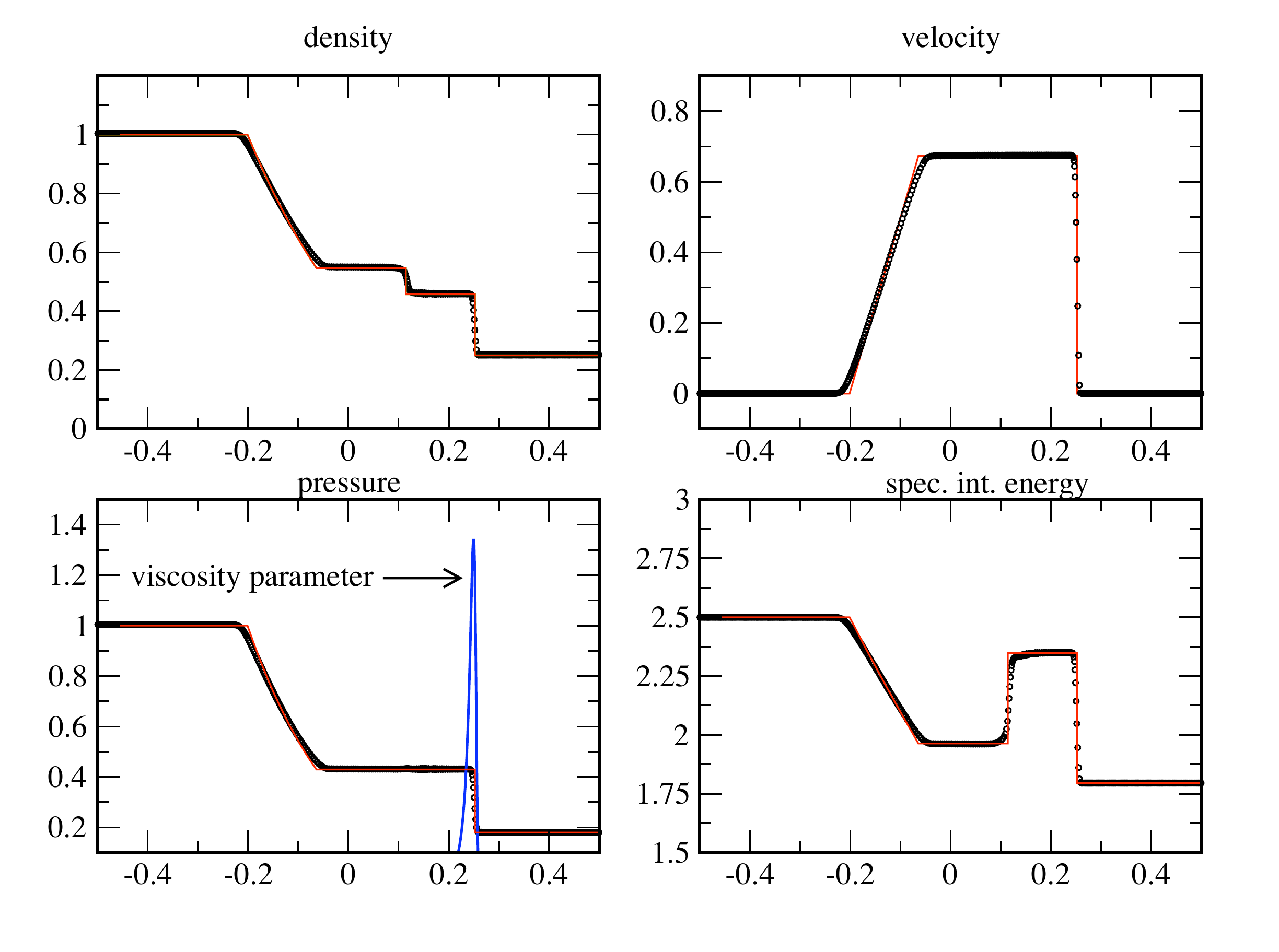} 
   \caption{Sod shock tube problem in 1D: the exact solution is given by the solid
  line, the numerical result is shown by the circles. In the lower left panel
  (pressure) the value of the time dependent artificial viscosity parameter
  $\alpha$ is overlaid. It is to be compared with the commonly used value
  $\alpha=1={\rm const}$.}
   \label{fig:sod}
\end{figure}

Due to the highly variable geometry and the importance of the strict
numerical conservation of physically conserved quantities, we use the smoothed
particle hydrodynamics method (SPH) to discretize the equations of an ideal
fluid. Using the SPH approximations\cite{benz90a,monaghan05} the conservation
of mass, momentum and energy translate into
\bea
\rho_a&=& \sum_b m_b W_{ab}\label{eq:basic:summary_sum_rho}\\
\frac{{d\vec{v}}_{a}}{dt}&=& - \sum_b m_b \left\{
\frac{P_a}{\rho_a^2} + \frac{P_b}{\rho_b^2}  + \Pi_{ab}
 \right\}\nabla_a W_{ab} + \vec{f}_{a,\rm
   grav}\label{eq:momentum_equation}\\ 
\frac{d u_{a}}{dt}&=& \frac{P_a}{\rho_a^2} \sum_b m_b \vec{v}_{ab} 
\nabla_a W_{ab} + \frac{1}{2} \sum_b m_b \Pi_{ab} \vec{v}_{ab}
\nabla_a W_{ab} \nonumber \\
& &+ \dot{\epsilon}_{{\rm nuc},a} \label{eq:energy_equation}. 
\eea
Here $\rho_a$ is the density at the position of particle $a$, $m_b$ the (constant) mass
of particle $b$, $W_{ab}= W(|\vec{r}_a-\vec{r}_b|, h_{ab})$
the cubic spline kernel\cite{monaghan85} with compact support whose width is set by the average of
the smoothing lengths of particle $a$ and $b$, $h_{ab}=(h_a+h_b)/2$. $P_a$ 
refers to the gas pressure, $\vec{f}_{a,\rm grav}$ to the gravitational
acceleration of particle $a$ and $\Pi_{ab}$ is the artificial 
viscosity tensor, see below, and $\vec{v}_{ab}= \vec{v}_a-\vec{v}_b$ with
$\vec{v}$ being the particle velocity. The quantity $u_a$ is the specific
internal energy of particle $a$, $\dot \epsilon_{{\rm nuc},a}$ is the thermonuclear energy  
generation, calculated in a operator split fashion as described below.   
Since SPH is a Lagrangian method, and we ignore mixing, the  
compositional evolution is a local phenomenon.\\
In cases where the geometry of the gas distribution varies substantially, it is
advisable to adapt the local resolution, i.e. the smoothing length, according
to the changes in the matter density. We achieve this by scaling the
smoothing length with the density according to
\be
\frac{h_a(t)}{h_{a,0}}= \left(\frac{\rho_{a,0}}{\rho_a(t)}\right)^{1/3}, \label{eq:ha_ha0}
\ee 
where the index $0$ labels the quantities at the beginning of the
simulation. The initial smoothing lengths are chosen so that each particle has
100 neighbors\footnote{A ``neighbor'' is a particle $b$ that yields a non-zero
  contribution to sums in Eqs.~(\ref{eq:basic:summary_sum_rho})-(\ref{eq:energy_equation}).}. 
Taking the Lagrangian time derivative of both sides of
Eq.~(\ref{eq:ha_ha0}) and using the Lagrangian form of the continuity equation,
one finds an evolution equation for the smoothing length of each particle:
\be
\frac{d h_a}{dt}= \frac{1}{3} h_a (\nabla \cdot \vec{v})_a,\label{eq:h_t}
\ee
where
\be 
(\nabla \cdot \vec{v})_a= - \frac{1}{\rho_a} \sum_b m_b \vec{v}_{ab}
\nabla_a W_{ab}.
\ee
This equation is integrated together with the other ODEs required for
hydrodynamics, Eqs.~(\ref{eq:momentum_equation}), (\ref{eq:energy_equation}),
(\ref{eq:alpha_t}), and possible changes in the abundances in the case of
nuclear reactions. During the integration the resulting new
neighbor number is constantly monitored for each particle and, if necessary,
an iteration of the smoothing length is performed to keep the neighbor
number in the desired range of between 80 and 120.\\
The SPH equations derived from a Lagrangian\cite{springel02,monaghan02} yield
different symmetries in the particle indices and additional multiplicative
factors with values close to unity, but a recent comparison\cite{rosswog07c}
between these two sets of equations showed only very minor differences in
practical applications. \\
We use the artificial viscosity tensor in the form given in \cite{monaghan92},
but with the following important modifications: i) the viscosity parameters
$\alpha$ and $\beta$ (commonly set to constants $\alpha=1$, $\beta=2$) are
replaced by $\alpha \rightarrow \tilde{\alpha}_{ab}= (\alpha_a + \alpha_b)
(f_a+f_b)/4$, $\beta \rightarrow 2 \tilde{\alpha}_{ab}$, where $f_k$ is the
so-called Balsara-switch\cite{balsara95} to suppress spurious forces in pure
shear flows, and ii) the viscosity parameter $\alpha_k$ is determined by
evolving an additional equation\cite{morris97} with a decay and a source
term\cite{rosswog00}: 
\be
\frac{d \alpha_{a}}{dt}= - \frac{\alpha_{a}- \alpha_{\rm min}}{\tau_{a}} +
S_{a} 
\ee
with
\be
 S_{a}= {\rm max}\left[-( \vec{\nabla} \cdot \vec{v})_{a}
  (\alpha_{\rm max}-\alpha_{a}),0\right]\label{eq:alpha_t}.
\ee
In the absence of shocks $\alpha$ decays to $\alpha_{\rm min}$ on a time scale
$\tau_a= h_{a}/(0.1 \; c_{a})$, where $c_a$ is the sound velocity. In a shock
$\alpha$ rises rapidly to capture the shock properly. A further discussion of
this form of artificial viscosity can be found in \cite{rosswog02a}.\\
To demonstrate the
ability of this scheme to capture shocks without spurious post-shock
oscillations we show in Fig.~\ref{fig:sod} the results of a standard,
one-dimensional Sod shock tube test\cite{sod78}  (with a polytropic gas of
adiabatic exponent $\Gamma= 1.4$) at t=0.17. For this test we used 1000 SPH
particles, $\alpha_{\rm max}= 1.5$ and $\alpha_{\rm min}= 0.1$. The numerical
solution (circles) agrees excellently with the exact solution (solid line),
the shock itself is spread over about 10 particles. In the third panel
(``pressure'') we have overlaid the value of the viscosity parameter
$\alpha$. It deviates substantially from its minimum value only in the direct
vicinity of the shock where it reaches values of about 1.3. This
is to be compared to the ``standard'' values $\alpha=1$ and $\beta=2$ that are
commonly applied in astrophysical simulations to each particle irrespective of
the necessity to resolve a shock. This scheme has been
shown\cite{rosswog00,rosswog02a} to  minimize possible artifacts due
to artificial viscosity.\\
Note that no efforts have been made to include the effects of heat transport.
Therefore, deflagration-type combustion cannot be handled adequately with the
current code. As will be discussed below, deflagration is of no importance for
the disruption processes that we discuss here.

\subsection{Equation of state}
The system of fluid equations (\ref{eq:basic:summary_sum_rho}),
(\ref{eq:momentum_equation}) and (\ref{eq:energy_equation}) needs to be closed
by an equation of state (EOS) that is appropriate for white dwarf matter. We
use the 
HELMHOLTZ EOS developed by the Center for Astrophysical Thermonuclear Flashes
at the University of Chicago. It accepts an externally calculated nuclear
composition which facilitates the coupling to reaction networks. 
The ions are treated as a Maxwell-Boltzmann gas, for the electron/positron gas
the exact expressions are integrated numerically (i.e. no assumptions about
the degree of degeneracy or relativity are made) and the result is
stored in a table. A sophisticated, biquintic 
Hermite polynomial interpolation is used to enforce the thermodynamic
consistency at interpolated values\cite{timmes00a}. The photon contribution is
treated as blackbody radiation.  
The EOS covers the density range form $10^{-10} \le \rho Y_e \le 10^{11}$ g
cm$^{-3}$ ($Y_e$ being the electron fraction\footnote{In the presence of
  electron-positron pairs it is given by $Y_e= \frac{n_{e^-} - n _{e^+}}
{\rho N_{\rm A}}$, where $n_{e^-}$/$n_{e^+}$ are the number densities of electrons/positrons.}) and temperatures from $10^4$ to
$10^{11}$ K.   

\subsection{Gravity}
The self-gravity of the fluid is calculated via a parallel version
of the binary tree described in \cite{benz90b}. The same tree is used to
search for the neighbor particles that are required for the density estimate,
Eq.~(\ref{eq:basic:summary_sum_rho}), and the gradients in
Eqs.~(\ref{eq:momentum_equation}), (\ref{eq:energy_equation}) and (\ref{eq:h_t}).
The gas acceleration due to a (Schwarzschild) black hole is treated in
the  Paczy\'nski-Wiita approximation\cite{pac80}. This approach has been
shown\cite{artemova96} to yield accurate results for the accretion onto
non-rotating black holes. To avoid numerical problems due to the singularity
at the Schwarzschild radius the pseudo potential is smoothly extended in a
non-singular way down to the hole\cite{rosswog05a} with an absorbing boundary
placed at a distance of $3 G M_{\rm bh}/c^2$ from the black hole with mass
$M_{\rm bh}$. 

\subsection{A minimal nuclear reaction network}
\begin{figure}
\centerline{
\includegraphics[width=8.5cm]{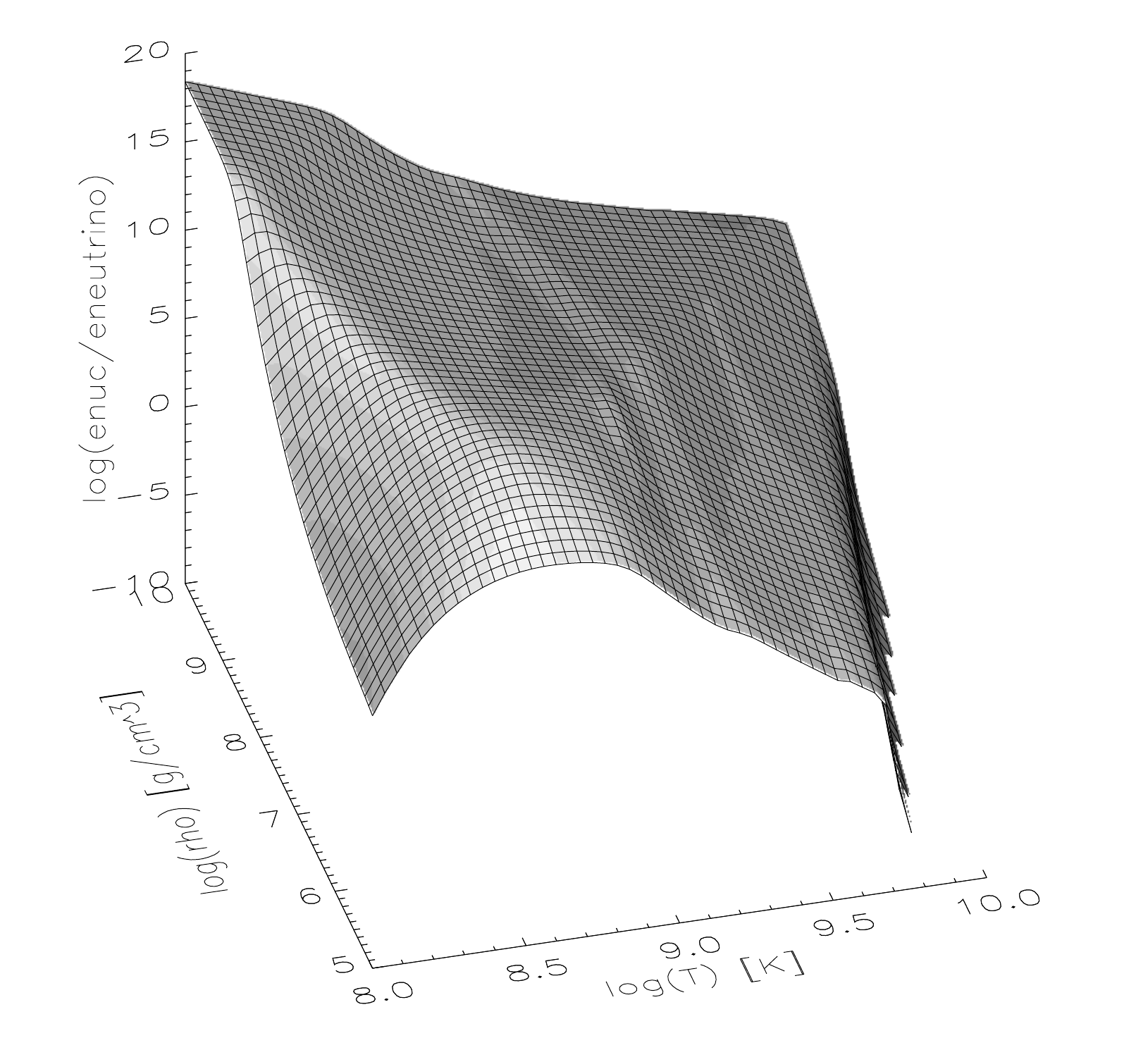}
\includegraphics[width=8cm]{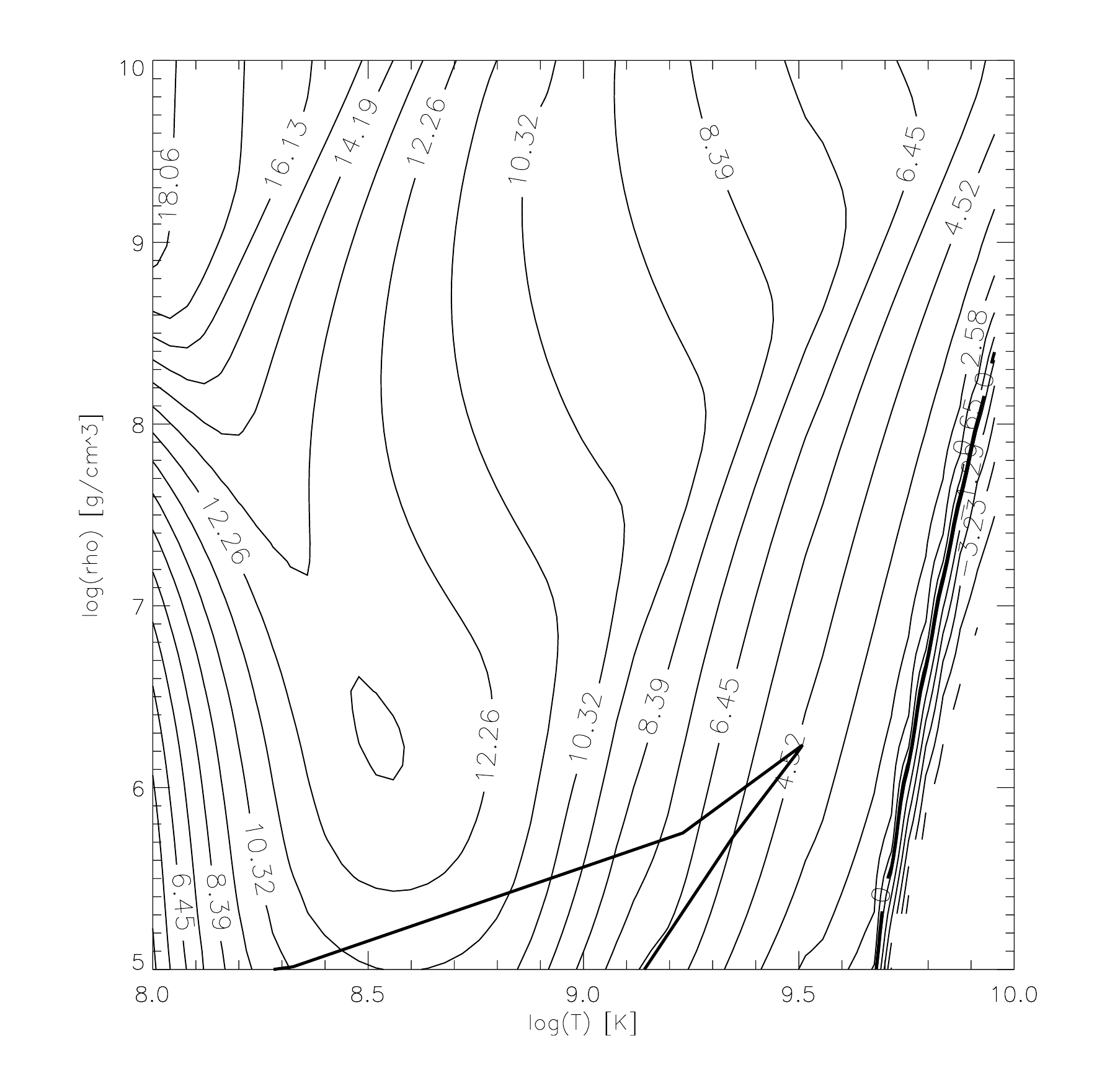}
}
\caption{Comparison of nuclear energy produced (from an initial pure He
  composition) and the energy lost via various neutrino reactions during
  1 s (the typical pericentre passage time, see Fig.~\ref{SMBH_WD_evo}).
  The thick solid line that starts at about $\log(T)= 9.1$ and ends at
  about $\log(T)= 8.3$ shows the trajectory of
  the hottest 10\% of the particles of the simulation shown in 
  Fig.~\ref{SMBH_WD_evo}). To see clearly where neutrino emission becomes
  dominant, we have also thickened the contour where nuclear and neutrino
  contributions are equal.}
\label{fig:enuc_vs_enu}
\end{figure}

To address whether tidal compression can trigger a thermonuclear
explosion, we need to evolve the nuclear composition and to
include the feedback onto the gas from the energy released by nuclear
burning. Running a full nuclear network with hundreds of species
for each SPH particle would be computationally prohibitive, therefore a 
``minimal'' network designed to provide the accurate energy 
generation\cite{hix98,timmes00b} is used. It
couples a conventional $\alpha$-network stretching from He to Si with a
quasi-equilibrium-reduced $\alpha$-network. The QSE-reduced network  
neglects reactions within small equilibrium groups that form at  
temperatures above 3.5 GK to reduce the number of abundance variables  
needed. Although a set of only 
seven nuclear species is used, this network reproduces all burning
stages from He-burning to nuclear statistical equilibrium  accurately. For
more details and tests we refer to \cite{hix98}.\\
In the presence of nuclear reactions the energy produced (or consumed) 
by nuclear reactions is given by
\be
\dot{\epsilon}_{\rm nuc,a}= N_{\rm A} \sum_j B_j \frac{dY_{j,a}}{dt},
\ee
where  $N_{\rm A}$ is Avogadro's number, $B_j$ 
is the nuclear binding energy of the nucleus $j$, $Y_{j,a}= n_{j,a}/(\rho_a
N_{\rm A})$ its abundance and $n_{j,a}$ is the number density of species
$j$. Again, the subscript $a$ indicates that these quantities are evaluated at
the position of particle $a$.\\
Since the nuclear reaction and the hydrodynamic time scales can differ by
many orders of magnitude, the network is coupled to the hydrodynamics in an
operator splitting fashion. In a first step,
Eqs.~(\ref{eq:momentum_equation}),(\ref{eq:energy_equation}),(\ref{eq:h_t}),(\ref{eq:alpha_t})
are integrated forward in time via a MacCormack predictor-corrector
scheme\cite{lomax01} with individual time steps\cite{rosswog07c} to obtain
new quantities at time $t^{n+1}$. In this step we ignore the nuclear source
term in Eq.~(\ref{eq:energy_equation}), the result is denoted by
$\tilde{u}_a^{n+1}$. This value has to be corrected for the energy release
that occurred from $t^{n}$ to $t^{n+1}$:
\begin{eqnarray}
\hspace*{-0.5cm}\epsilon_{a,n\rightarrow n+1}&=& N_{\rm A}\hspace*{-0.1cm} \sum_j\hspace*{-0.1cm} B_j \hspace*{-0.1cm}
\int_{t^{n}}^{t^{n+1}} \hspace*{-0.4cm}\frac{dY_{j,a}}{dt} (\rho_a(t),T_a(t),Y_{k,a}(t))  dt\\
&=& N_A \sum_{j=1} B_j (Y_{j,a}^{n+1}-Y_{j,a}^n),
\end{eqnarray}
where $\rho_a(t) \approx \rho_a(t^n) + \frac{t-t^n}{t^{n+1}-t^n}
\{\rho_a(t^{n+1})-\rho_a(t^n)\}$ and $T_a(t)\approx T_a(t^n)$ has been used to
integrate the abundances $Y_{j,a}$ via the implicit backward Euler method
(the network integration is described in detail in\cite{hix98}). The final
value for the specific energy at time $t^{n+1}$ is given by
\be
u_a^{n+1}= \tilde{u}_a^{n+1} + \epsilon_{a,n\rightarrow n+1}.
\ee
Now the EOS is called again to make all thermodynamic quantities 
consistent with this new value $u_a^{n+1}$. Once the derivatives have 
been updated, the procedure can be repeated for the next time step.\\ 
For the hydrodynamic time step we use the minimum of several criteria. We use
a force criterion and a combination of Courant-type and viscosity-based
criterion\cite{monaghan92} 
\be
\Delta t_{f,a}= \sqrt{h_a/|\vec{f}_a|}
\ee
and
\be 
\Delta t_{C,a}= \frac{h_a}{v_{{\rm s},a} + 0.6(v_{{\rm s},a}+2 \; {\rm max}_b
  \; \mu_{ab})},
\ee
where $\vec{f}_a$ is the acceleration, $v_{{\rm s},a}$ is the sound velocity
and $\mu_{ab}$ a quantity used in the artificial viscosity tensor (for its
explicit form see \cite{monaghan92}). To ensure a close coupling between
hydrodynamics and nuclear reactions in regions where burning
is expected, we apply two additional time step criteria. One triggers on matter
compression, the other on the distance to the black hole:
\be
\Delta t_{\rm comp,a}= - 0.03/(\nabla \cdot \vec{v})_a
\ee
and
\be
\Delta t_{\rm bh,a}=     0.03/\sqrt{G M_{\rm bh}/r_{\rm bh,a}^3}.
\ee 
The ``desired'' hydro time step of each particle is then chosen as 
\be
\Delta t_{{\rm des},a}= 0.2 \; {\rm min}(\Delta t_{f,a}, \Delta t_{C,a},
\Delta t_{\rm comp,a}, \Delta t_{\rm bh,a}).
\ee
How these desired time steps are
transformed into the individual block time steps is explained in detail
in\cite{rosswog07c}. As a test, low-resolution versions of the production runs
were run once with this time step prescription and once with halved
time steps. No noticeable differences could be found.\\
One may wonder whether neutrino emission could subduct substantial amounts of
energy during the burning phase. To test for this, we have plotted in 
Fig.~\ref{fig:enuc_vs_enu} the ratio of the nuclear energy produced and the
energy lost to neutrinos within a typical pericentre passage of 1 s
duration. For the neutrino emission pair annihilation, plasma, photoneutrino,
bremsstrahlung and recombination processes were considered according to the
fit formulae of \cite{itoh96} as coded by
F. Timmes\footnote{http://cococubed.asu.edu}. For the conditions 
relevant to this study neutrino emission was never relevant.  

\section{Results}
\label{results}
\begin{figure}
\centerline{
\includegraphics[width=9cm]{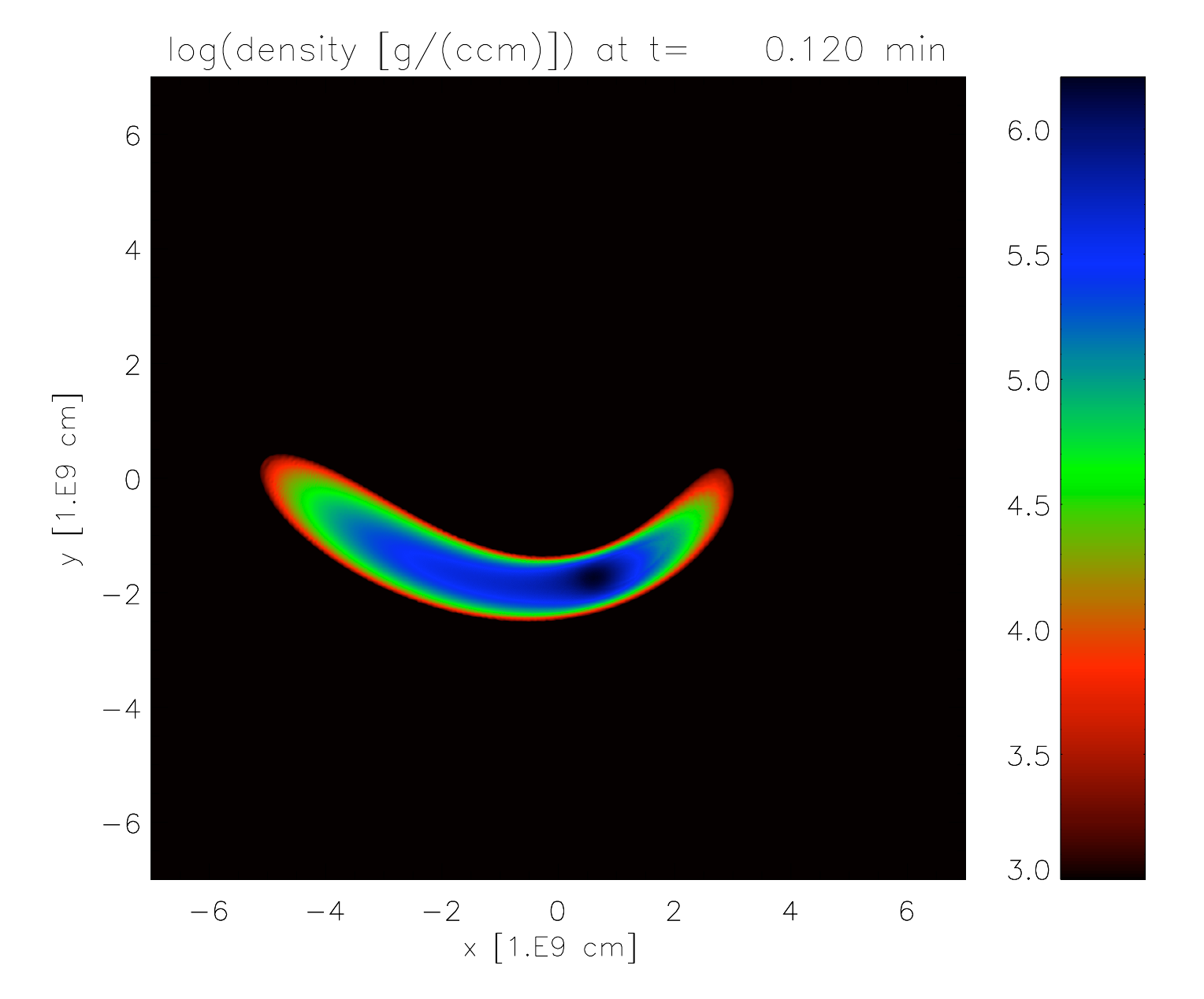}
\includegraphics[width=9cm]{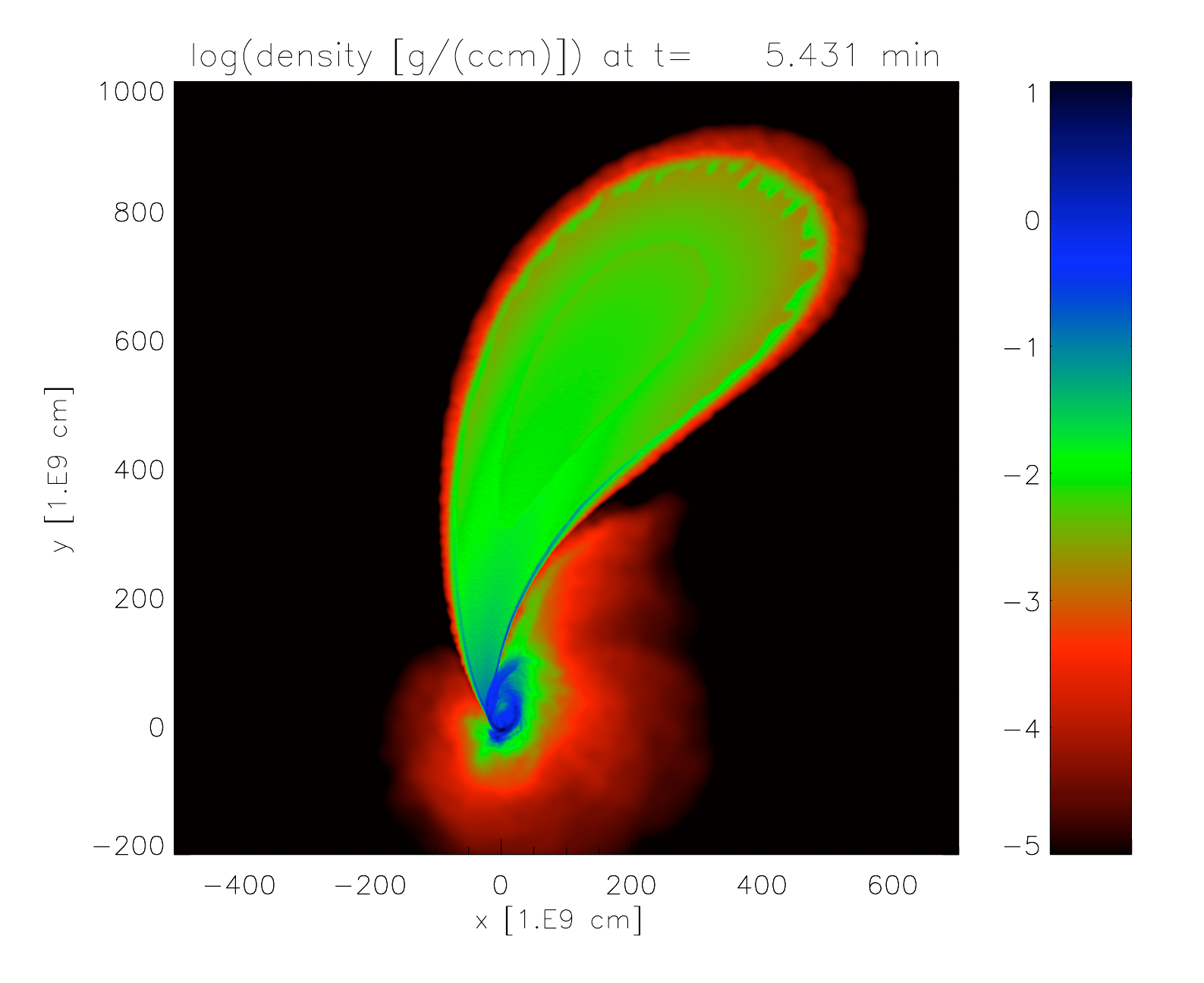}
}
\caption{Tidal disruption of a 0.2  M$_\odot$ He white dwarf by a 1000
  M$_\odot$ black hole (located at the origin).}
\label{SMBH_WD_evo}
\end{figure}

\begin{figure}
\centerline{\includegraphics[width=14cm]{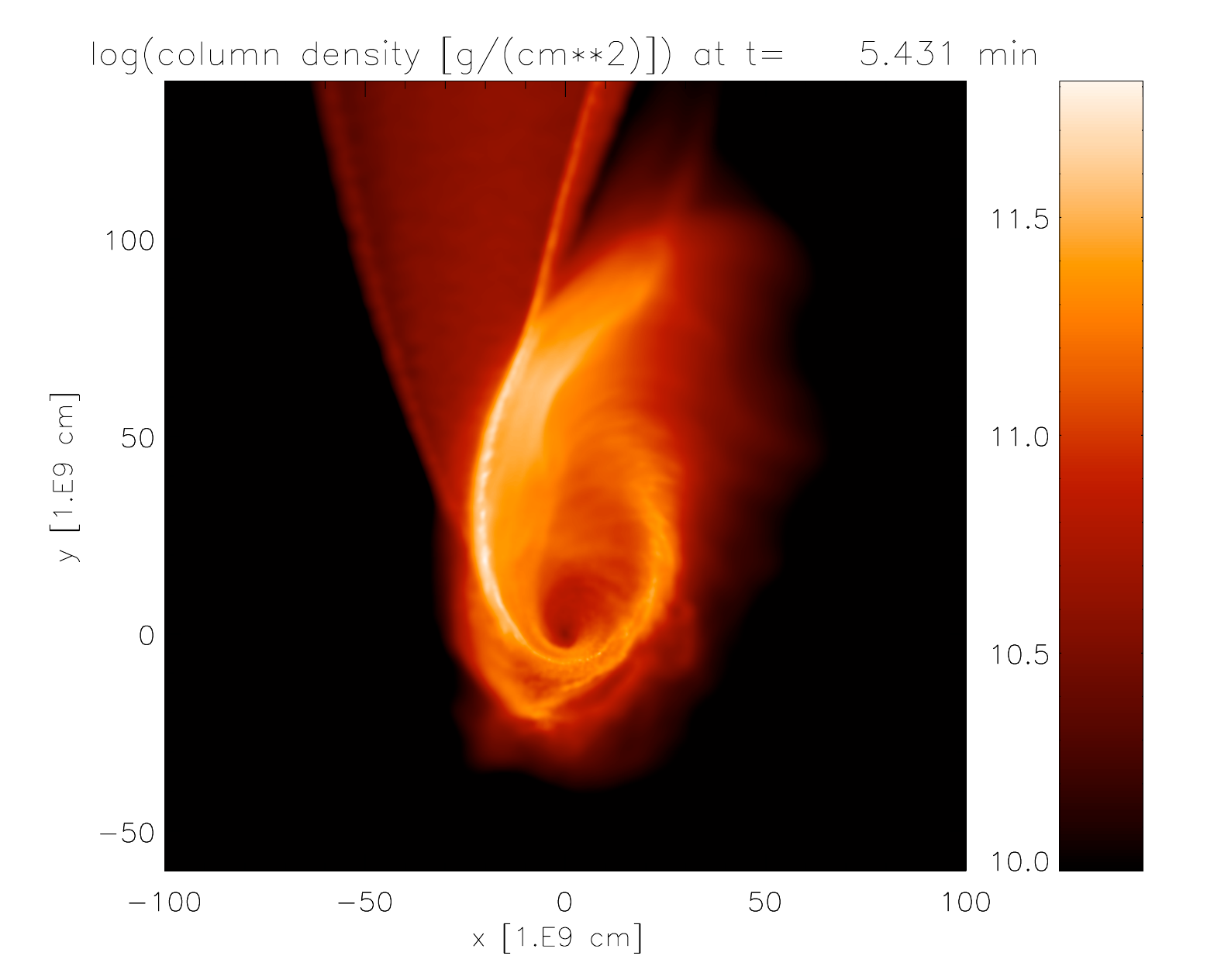}}
\caption{Zoom-in on the forming accretion disk. Color-coded is the column
 density (in g/cm$^2$), angular momentum is redistributed in the shock that
forms when the accretion stream interacts with itself.}
\label{SMBH_WD_ang_mom_shock}
\end{figure}

As initial condition we set up a white
dwarf on a parabolic orbit around the black hole with an initial separation of
several tidal radii $R_{\rm tid}= R_{\rm WD} \left( \frac{M_{\rm bh}}{M_{\rm
      WD}}\right)^{1/3}$. 
In a large set of simulations we explored black hole masses of 100,
500, 1000, 5000 and 10 000 \Msun and white dwarf masses of 0.2, 0.6 and 1.2
\msun. Each time several ``penetration factors'', $P= R_{\rm tid}/R_{\rm
  peri}$, where $R_{\rm peri}$ is the pericentre separation, were explored. To
be conservative, we set the initial  
white dwarfs to very low temperatures ($T_0= 5\cdot 10^4$ K). The numerical
resolution varied from 500 000 to more than $4 \cdot 10^6$ SPH particles.
In all cases we found explosions (nuclear energy release larger than the white
dwarf gravitational binding energy) whenever the penetration factors exceeded
values of about 3. Further details of these simulations will be discussed
elsewhere\cite{rosswog07d}.\\
For conciseness we focus here on one exemplary simulation of a 0.2 \msun, 
pure He white dwarf (modeled with more than $4\cdot10^6$ SPH particles) and a 
1000 \Msun black hole, see Fig.~\ref{SMBH_WD_evo}. The
first snapshot (0.12 minutes after the simulation start) shows the stage of
maximum white dwarf compression at pericentre passage, in which the white
dwarf is also severely compressed perpendicular to the orbital plane. The peak
compression occurs at a spatially fixed point (seen as the density peak in
Fig.~\ref{SMBH_WD_evo}, left). The white dwarf fuel is fed with free-fall
velocity $v_{\rm ff}= (2 G M_{\rm bh}/R_{\rm peri})^{1/2}
= 1.6 \cdot 10^{5} {\rm km \; s}^{-1} \; \left(M_{\rm bh}/1000 
{\rm M}_{\odot}\right)^{1/2} \left(R_{\rm peri}/10^4 {\rm km}\right)^{-1/2}$ 
into this compression point. The comparison with typical flame propagation
speeds ($\sim 100$ km/s) shows that combustion effects can be safely
neglected for this investigation.\\ 
During
the short compression time (of order one second), the peak density increases by
more than one order of magnitude (with respect to the initial, unperturbed
star) to $>6\cdot 10^5$ \gcc, the peak temperatures get close to nuclear
statistical equilibrium ($>3.9\cdot 10^9$ K). During this stage 0.11 \Msun are
burnt, mainly into silicon group (74\%), iron group (22.5\%) elements and
carbon. The nuclear energy release triggers a thermonuclear explosion of 
the white dwarf. Since a much smaller fraction ends up in iron group nuclei 
(nickel), this explosion is underluminous in comparison to a normal type I
a supernova which produces $\sim 0.5$ \Msun of nickel. Due to the very
different geometry, the lightcurves of such explosions are expected to
deviate substantially from standard type Ia light curves. This topic deserves
further detailed investigations.\\  
About 35 \% of the initial stellar mass remain gravitationally bound to the
black hole and will subsequently by accreted. During infall, matter
trajectories become radially focused towards the pericentre. The large spread
in the specific energy across the accretion stream width produces a large
spread of apocentric distances and thus a fan-like spraying of the white dwarf
debris after pericentre passage. This material interacts with the infalling
material in an angular momentum redistribution shock, see
Fig.~\ref{SMBH_WD_ang_mom_shock}, which results in the circularization 
of the forming accretion disk. The subsequent accretion onto the black hole
produces a soft X-ray flare close the Eddington-luminosity for a duration of
months.\\ 
For other combinations of black hole and white dwarf masses with $P > 3$ 
we found a similar behavior. Therefore, an underlumious thermonuclear explosion
accompanied by soft X-ray flare may whistle-blow the existence of intermediate
mass black holes in globular clusters. The rate of this particular type of
explosion amounts to a few tenths of a percent of ``standard'' type Ia
supernovae. Future supernova surveys such as SNF could be able to detect 
several of these events. \\

\noindent{\bf Acknowledgments} \\
We thank Holger Baumgardt, Peter Goldreich, Jim Gunn,
Piet Hut, Dan Kasen and Martin Rees for very useful
discussions. The simulations presented in this paper were performed on the JUMP
computer of the H\"ochstleistungsrechenzentrum J\"ulich.
E. R. acknowledges support from the DOE Program for
Scientific Discovery through Advanced Computing (SciDAC;
DE-FC02-01ER41176). W. R. H. has been partly supported by
the National Science Foundation under contracts PHY-0244783 and AST-0653376.
Oak Ridge National Laboratory is managed by UT-Battelle, LLC, for the  
U.S. Department of Energy under contract DE-AC05-00OR22725.


\clearpage

\clearpage

\clearpage

\clearpage


\begin{thebibliography}{}
\bibitem{artemova96}
I.~V. {Artemova}, G.~{Bjoernsson}, I.~D. {Novikov}, {Modified Newtonian
  Potentials for the Description of Relativistic Effects in Accretion Disks
  around Black Holes}, ApJ 461 (1996) 565.

\bibitem{balsara95}
D.~Balsara, von neumann stability analysis of smooth particle
  hydrodynamics--suggestions for optimal algorithms, J. Comput. Phys. 121
  (1995) 357.

\bibitem{benz90a}
W.~Benz, Smooth particle hydrodynamics: A review, in: J.~Buchler (ed.),
  Numerical Modeling of Stellar Pulsations, Kluwer Academic Publishers,
  Dordrecht, 1990, p. 269.

\bibitem{benz90b}
W.~Benz, R.~Bowers, A.~Cameron, W.~Press, ApJ 348 (1990) 647.

\bibitem{dearborn05}
D.~S.~P. {Dearborn}, J.~R. {Wilson}, G.~J. {Mathews}, {Relativistically
  Compressed Exploding White Dwarf Model for Sagittarius A East}, ApJ 630
  (2005) 309--320.

\bibitem{gebhardt02}
K.~{Gebhardt}, R.~M. {Rich}, L.~C. {Ho}, {A 20,000 M$_\odot$ Black Hole in the
  Stellar Cluster G1}, ApJL 578 (2002) L41--L45.

\bibitem{gebhardt05}
K.~{Gebhardt}, R.~M. {Rich}, L.~C. {Ho}, {An Intermediate-Mass Black Hole in
  the Globular Cluster G1: Improved Significance from New Keck and Hubble Space
  Telescope Observations}, ApJ 634 (2005) 1093--1102.

\bibitem{gerssen02}
J.~{Gerssen}, R.~P. {van der Marel}, K.~{Gebhardt}, P.~{Guhathakurta}, R.~C.
  {Peterson}, C.~{Pryor}, {Hubble Space Telescope Evidence for an
  Intermediate-Mass Black Hole in the Globular Cluster M15. II. Kinematic
  Analysis and Dynamical Modeling}, AJ 124 (2002) 3270--3288.

\bibitem{gerssen03}
J.~{Gerssen}, R.~P. {van der Marel}, K.~{Gebhardt}, P.~{Guhathakurta}, R.~C.
  {Peterson}, C.~{Pryor}, {Addendum: Hubble Space Telescope Evidence for an
  Intermediate-Mass Black Hole in the Globular Cluster M15. II. Kinematic
  Analysis and Dynamical Modeling}, AJ 125 (2003) 376--377.

\bibitem{hix98}
W.~R. {Hix}, A.~M. {Khokhlov}, J.~C. {Wheeler}, F.-K. {Thielemann}, {The
  Quasi-Equilibrium-reduced alpha -Network}, ApJ 503 (1998) 332--+.

\bibitem{itoh96}
N.~Itoh, H.~Hayashi, A.~Nishikawa, Y.~Kohyama, ApJ 339 (1989) 354.

\bibitem{lewin06}
W.~H.~G. {Lewin}, M.~{van der Klis}, {Compact stellar X-ray sources}, Compact
  stellar X-ray sources, 2006.

\bibitem{lomax01}
H.~Lomax, T.~Pulliam, D.~Zingg, Fundamentals of Computational Fluid Dynamics,
  Springer, Berlin, 2001.

\bibitem{luminet89b}
J.-P. {Luminet}, B.~{Pichon}, {Tidal pinching of white dwarfs}, A\&A 209 (1989) 103--110.

\bibitem{monaghan85}
J.~Monaghan, J.~Lattanzio, A refined particle method for astrophysical
  problems, A\&A 149 (1985) 135.

\bibitem{monaghan92}
J.~J. Monaghan, Ann. Rev. Astron. Astrophys. 30 (1992) 543.

\bibitem{monaghan02}
J.~J. {Monaghan}, {SPH compressible turbulence}, MNRAS 335 (2002) 843--852.

\bibitem{monaghan05}
J.~J. {Monaghan}, {Smoothed particle hydrodynamics}, Reports of Progress in
  Physics 68 (2005) 1703--1759.

\bibitem{morris97}
J.~Morris, J.~Monaghan, A switch to reduce sph viscosity, J. Comp. Phys. 136
  (1997) 41.

\bibitem{pac80}
B.~{Paczynsky}, P.~J. {Wiita}, {Thick accretion disks and supercritical
  luminosities}, A\&A 88 (1980) 23--31.

\bibitem{pooley06}
D.~{Pooley}, S.~{Rappaport}, {X-Rays from the Globular Cluster G1:
  Intermediate-Mass Black Hole or Low-Mass X-Ray Binary?}, ApJL 644 (2006)
  L45--L48.

\bibitem{portegies04}
S.~F. {Portegies Zwart}, H.~{Baumgardt}, P.~{Hut}, J.~{Makino}, S.~L.~W.
  {McMillan}, {Formation of massive black holes through runaway collisions in
  dense young star clusters}, Nature 428 (2004) 724--726.

\bibitem{richstone98}
D.~{Richstone}, E.~A. {Ajhar}, R.~{Bender}, G.~{Bower}, A.~{Dressler}, S.~M.
  {Faber}, A.~V. {Filippenko}, K.~{Gebhardt}, R.~{Green}, L.~C. {Ho},
  J.~{Kormendy}, T.~R. {Lauer}, J.~{Magorrian}, S.~{Tremaine}, {Supermassive
  black holes and the evolution of galaxies.}, Nature 395 (1998) A14+.

\bibitem{rosswog05a}
S.~{Rosswog}, {Mergers of Neutron Star-Black Hole Binaries with Small Mass
  Ratios: Nucleosynthesis, Gamma-Ray Bursts, and Electromagnetic Transients},
  ApJ 634 (2005) 1202--1213.

\bibitem{rosswog02a}
S.~{Rosswog}, M.~B. {Davies}, {High-resolution calculations of merging neutron
  stars - I. Model description and hydrodynamic evolution}, MNRAS 334 (2002)
  481--497.

\bibitem{rosswog00}
S.~{Rosswog}, M.~B. {Davies}, F.-K. {Thielemann}, T.~{Piran}, {Merging neutron
  stars: asymmetric systems}, A\&A 360 (2000) 171--184.

\bibitem{rosswog07f}
S.~Rosswog, E.Ramirez-Ruiz, R.~Hix, Atypical thermonuclear supernovae from
  tidally crushed white dwarfs, ApJ, accepted, arXiv:0712.2513.

\bibitem{rosswog07d}
S.~Rosswog, E.Ramirez-Ruiz, R.~Hix, 
to be submitted.

\bibitem{rosswog07c}
S.~Rosswog, D.~Price, Magma: a magnetohydrodynamics code for merger
  applications, MNRAS 379 (2007) 915 -- 931.

\bibitem{sod78}
G.~Sod, A survey of several finite difference methods for systems of nonlinear
  hyperbolic conservation laws, J. Comput. Phys. 43 (1978) 1--31.

\bibitem{springel02}
V.~{Springel}, L.~{Hernquist}, {Cosmological smoothed particle hydrodynamics
  simulations: the entropy equation}, MNRAS 333 (2002) 649--664.

\bibitem{timmes00b}
F.~X. {Timmes}, R.~D. {Hoffman}, S.~E. {Woosley}, {An Inexpensive Nuclear
  Energy Generation Network for Stellar Hydrodynamics}, ApJS 129 (2000)
  377--398.

\bibitem{timmes00a}
F.~X. {Timmes}, F.~D. {Swesty}, {The Accuracy, Consistency, and Speed of an
  Electron-Positron Equation of State Based on Table Interpolation of the
  Helmholtz Free Energy}, ApJS 126 (2000) 501--516.

\bibitem{wilson04}
J.~R. {Wilson}, G.~J. {Mathews}, {White Dwarfs near Black Holes: A New Paradigm
  for Type I Supernovae}, ApJ 610 (2004) 368--377.

\bibitem{zezas02}
A.~{Zezas}, G.~{Fabbiano}, A.~H. {Rots}, S.~S. {Murray}, {Chandra Observations
  of ``The Antennae'' Galaxies (NGC 4038/4039). III. X-Ray Properties and
  Multiwavelength Associations of the X-Ray Source Population}, ApJ 577 (2002)
  710--725.

\end{thebibliography}
\end{document}